\documentstyle[preprint,aps,epsf,floats,psfig]{revtex}

\tighten

\newcommand{\beq}{\begin{equation}}
\newcommand{\eeq}{\end{equation}}
\newcommand{\beqa}{\begin{eqnarray}}
\newcommand{\eeqa}{\end{eqnarray}}

\newcommand{\ga}{g_A}
\newcommand{\gas}{g_A^2}

\newcommand{\mpi}{m_\pi}
\newcommand{\mpis}{m_\pi^2}

\newcommand{\lsim}{\raisebox{-0.7ex}{$\stackrel{\textstyle <}{\sim}$ }}

\def\si{{}^1\kern-.14em S_0}
\def\siii{{}^3\kern-.14em S_1}
\def\piii{{}^3\kern-.14em P_1}
\def\diii{{}^3\kern-.14em D_1}

\begin{document}

\begin{titlepage}

\hfill{NT@UW-02-023}

\vspace{0.5cm}

\begin{center}
{\Large\bf The Quark-Mass Dependence of Two-Nucleon Systems}

\vspace{1.2cm}

{\bf Silas R.~Beane and Martin J.~Savage}

\vspace{0.2cm}
{\it
Department of Physics,
University of Washington,\\
Seattle, WA 98195}
\end{center}

\vspace{0.1cm}

\begin{abstract}
  
  We explore the quark-mass dependence of two-nucleon systems.  Allowed regions
  for the scattering lengths in the $\si$ and $\siii$ channels as functions of
  the light-quark masses are determined from the current uncertainty in
  strong-interaction parameters that appear at next-to-leading order in the
  effective field theory. Where experimental constraints are absent, as is the
  case for the quark-mass dependent four-nucleon operators, we use naive
  dimensional analysis. We find it likely that there is no bound state in the
  $\si$ channel in the chiral limit. However, given the present uncertainties
  in strong-interaction parameters it is unclear whether the deuteron is bound
  or unbound in the chiral limit.

\end{abstract}

\vspace{2cm}
\vfill
\end{titlepage}
\setcounter{page}{1}

%%%%%%%%%%  Intro %%%%%%%%%%%%%%%%
\section{Introduction}

An important aspect of strong-interaction physics that is only now being
explored is the dependence of nuclei and nuclear reactions on the fundamental
parameters of nature, such as $\Lambda_{\rm QCD}$ and the quark masses, $m_q$. 
With recent developments in the effective field theory (EFT) description of
multi-nucleon systems fundamental questions about nuclear physics can be
addressed.  While the $m_q$-dependence of the nuclear force is unrelated to
present day observables, it is a fundamental aspect of nuclear physics, and in
some sense serves as a benchmark for the development of a perturbative theory
of nuclear forces.  Having this behavior under control will be essential to any
bridge between lattice QCD simulations and nuclear physics in the near future.
In work by Bedaque, van Kolck and the authors~\cite{Beane:2001bc} (which we
will refer to as BBSvK), a new power-counting scheme, BBSvK-counting, was
introduced, and with a partial next-to-leading order (NLO) calculation in this
new counting scheme, an attempt was made to determine the behavior of the
two-nucleon force as the $m_q$ were increased or
decreased from their physical values.  Calculations of the $m_q$-dependence of
scattering in the $\si$-channel and $\siii-\diii$ coupled-channels as well as a
discussion of the deuteron binding energy were presented.  In a subsequent
paper~\cite{Beane:2002vq} we performed the first complete NLO calculation in
BBSvK-counting of the $m_q$-dependence of the two-nucleon systems.  Previous
estimates of such $m_q$-dependences~\cite{Bulgac:1997ji,timedep} that served as
input to place bounds on physics beyond the standard model via the
time-variation of fundamental parameters on cosmological time scales were found
to be incomplete.

In this work, which should be considered a sequel to Ref.~\cite{Beane:2002vq},
we investigate the $m_q$-dependence of the two-nucleon systems, incorporating the
present uncertainties in strong interaction parameters~\footnote{ We are
  grateful to J.~Bjorken who suggested this work to us during {\it Marshall
    Baker Symposium} held at the UW in June 2002. }.  In some instances,
parameters that play a central role in the $m_q$-dependence, such as the
leading $m_q$-dependent four-nucleon interaction, are not constrained
experimentally. In such cases, we use
naive-dimensional-analysis (NDA) to provide a ``reasonable range'' for their
values.  Of particular interest to us is the behavior of the two-nucleon sector
in the chiral limit.  We find it likely that there is no di-nucleon bound state
in the $\si$-channel; i.e. the scattering length remains negative.  On the
other hand, we cannot determine if there is a bound-state in the $\siii-\diii$
coupled channels in the chiral limit.  The strong interaction parameters are
sufficiently uncertain at present to preclude a definitive answer to this
fundamental question. 

%%%%%%%%%%%%%%%%
\section{The $\si$-Channel and the Di-Nucleon}

In the $\si$-channel, pions give a subleading contribution to the scattering
amplitude and are included in perturbation theory. In this channel,
BBSvK-counting is equivalent to Kaplan-Savage-Wise (KSW) power-counting~\cite{KSWb}.
Thus, analytic expressions for the scattering amplitude, and hence the scattering length, as a
function of $m_q$ and momentum, $p$, are straightforwardly found. From the
NLO amplitude~\cite{KSWb} it is easy to construct $p\cot\delta^{(\si)}$, which
has a well-behaved power-series expansion for $p < m_\pi/2$, and thus a linear
combination of $C_0$ (the coefficient of the $m_q$- and $p$-independent
four-nucleon operator) and $D_2$ (the coefficient of the $p$-independent and
leading $m_q$-dependent four-nucleon operators) can be determined in terms
of the scattering length $a^{(\si)}=-23.714\pm 0.013~{\rm fm}$ 
at the physical value of the pion mass.
Furthermore, $C_2$ (the coefficient of the $m_q$-independent, leading
$p^2$-dependent four-nucleon operator) can be determined in terms of the
effective range, $r^{(\si)}=2.73\pm 0.03~{\rm fm}$.  Once these parameters are
fixed, the scattering length, effective range and phase-shift can be determined
as a function of $m_q$ at NLO in the KSW expansion.  The scattering length 
is, in the isospin limit,
\begin{eqnarray}
\frac{1}{a^{(\si)}}=\gamma +\frac{g_A^2 M_N}{8\pi f_\pi^2}\left[ m_\pi^2\,
  \log{\left({\mu\over m_\pi}\right)}+ (\gamma - m_\pi )^2
-(\gamma - \mu )^2 \right] 
-\frac{M_N m_\pi^2}{4\pi}(\gamma -\mu )^2 \ D_2(\mu)
\ \ ,
\label{eq:scattlength}
\end{eqnarray}
where $\gamma =\mu +4\pi /M_N C_0(\mu)$, and $\mu$ is the PDS renormalization
scale~\cite{KSWb}.  We have used the leading order (LO) relation between the
pion mass, $m_\pi$, and the light-quark masses, $m_q$.  
As pions are treated in perturbation theory, 
the $m_q$-dependence arising from the chiral
expansions of $g_A$, $f_\pi$ and $M_N$ is formally higher order in
KSW-counting and  is neglected. 
The physical values of these parameters, 
$g_A\sim 1.25$, $f_\pi\sim 132~{\rm  MeV}$ and $M_N\sim 939~{\rm MeV}$,
are used for the analysis in this channel.

At present, the coefficient $D_2$ has not been experimentally separated from the
coefficient $C_0$ in this channel, or in the $\siii-\diii$ coupled-channels.
This is because both parameters are coefficients of momentum-independent
operators with the only difference between the operators arising
from their couplings to pions; the $C_0$ operator does not give rise to
interactions between two-nucleons and multiple pions, while the $D_2$ operator
does. In order to determine the range of values of
$D_2$ consistent with NDA, we use the
expression for the scattering length in eq.~(\ref{eq:scattlength}) at the
physical value of the pion mass and find values of $C_0 ( m_\pi^{{\rm PHYS}}
)$ and $D_2( m_\pi^{{\rm PHYS}} )$ that satisfy the constraint
\begin{eqnarray}
& & | (m_\pi^{\rm PHYS})^2 D_2 \  | \ <\ \eta\ | C_0 |
\ \ \ ,
\label{eq:etadef}
\end{eqnarray}
where the constant $\eta$ is chosen to be $1/5$ and $1/15$ for demonstrative 
purposes.  
\begin{figure}[ht]
\centerline{\psrotatefirst
\psfig{file=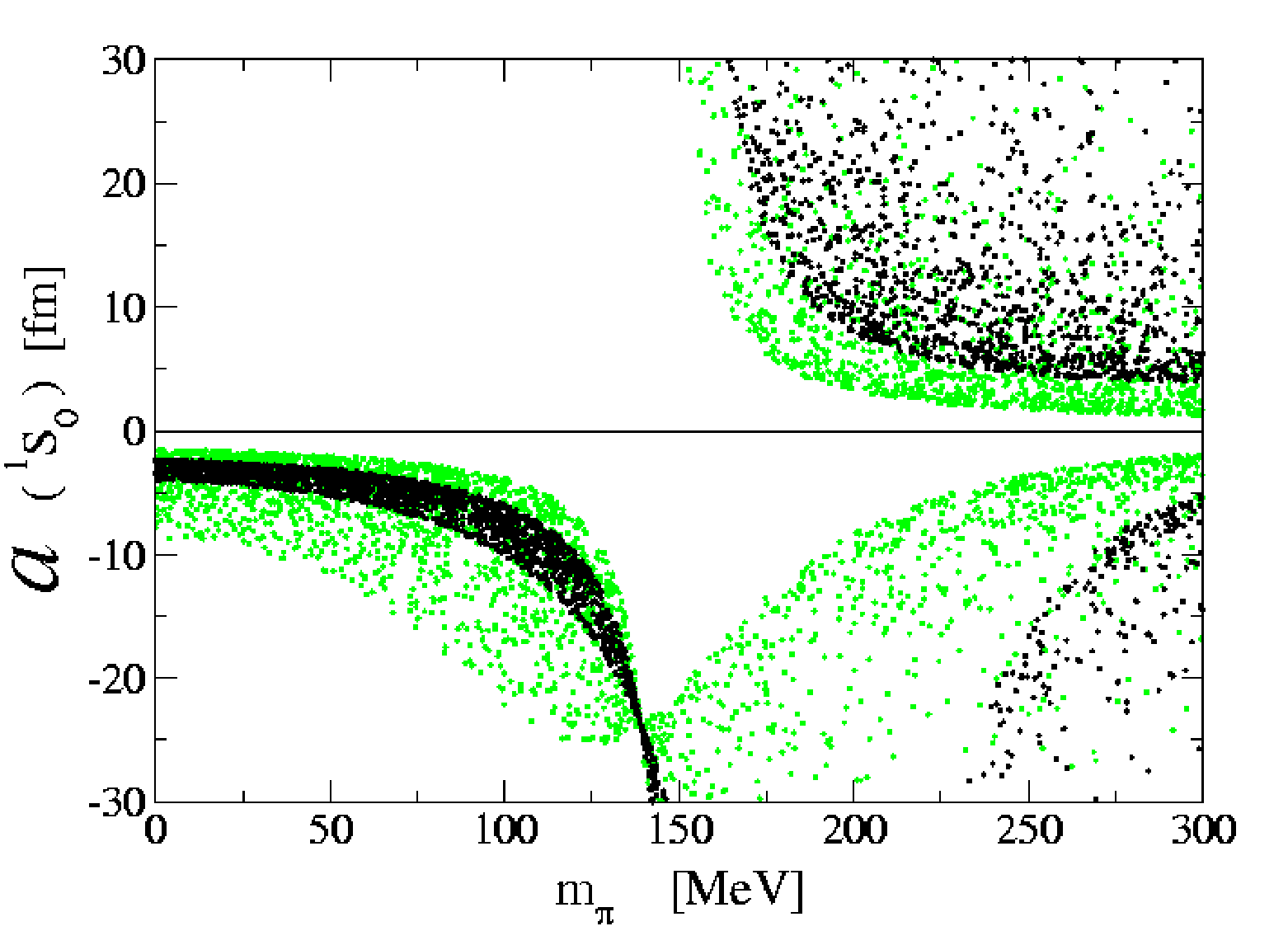,width=3.3in}}
\vskip 0.15in
\noindent
\caption{The scattering length in the $\si$-channel as a function of the 
pion mass.
The light gray region corresponds to $\eta=1/5$ and the 
black region corresponds to $\eta=1/15$.
At the physical value of the pion mass the scattering length is
$a^{(\si)}=-23.714\pm 0.013~{\rm fm}$.
}
\label{fig:1S0mq}
\vskip .2in
\end{figure}
It is worthwhile
discussing why these values are chosen.  Very roughly, 
$\eta\sim (m_\pi /\Lambda)^2$ where $\Lambda$ represents the scale 
at which the EFT description breaks down. 
Naively, the KSW perturbative-pion expansion breaks down at a scale 
$\Lambda_{\rm NN}=8\pi f_\pi^2/\ga^2 M\sim 300~{\rm MeV}$. This would
then give $\eta\sim 1/5$. On the other hand, one might expect that a better
estimate of the cutoff scale is given by the mass of the fictitious
scalar, $\sigma$, that provides intermediate-range attraction in many
potential models. Taking $m_\sigma\sim 500-600~{\rm MeV}$ one then arrives
at $\eta\sim 1/15$. Of course this last argument is somewhat specious;
it is really at $m_\sigma /2\sim \Lambda_{\rm NN}$ that $\sigma$
should be treated as a dynamical object. This discussion underscores the
difficulty in providing an {\it a priori} estimate of a parameter in
an EFT. Hence, given the complete lack of experimental information on $D_2$,
and the absence of any reliable method to estimate its value, we take 
$\eta$ to be $1/5$ and $1/15$ to provide an estimate of the uncertainty 
arising from $D_2$.  
We stress that even larger values of $\eta$, such as $1/3$, are not
excluded experimentally. Of course larger values of $\eta$ lead to
a smaller radius of convergence of the EFT.
In Fig.~\ref{fig:1S0mq} we show the scattering
length in the $\si$-channel as a function of $m_q$, where we have
randomly sampled values of $m_\pi$ and $\eta$ within the allowed
ranges\footnote{We choose to use scatter plots as the point density
reflects the probability associated with a particular set of
low-energy constants.}. From the standpoint of NDA, it is likely that the 
di-neutron is unbound in the chiral limit for any values of $m_q$ less than the
physical values.
Of course, the same will be true for the other di-nucleon states 
in the $\si$-channel by isospin symmetry.
On the other hand, for $m_q$ larger than their physical values, we see that
there are both bound and unbound di-neutrons for values of the 
strong-interaction parameters consistent with NDA.
This observation reinforces our previous analysis~\cite{Beane:2002vq}.

For the strong-interaction parameter sets where the scattering length remains
unnaturally large in the chiral limit, it is likely that higher-order
contributions to the scattering amplitude will significantly disturb the 
system.
Thus, for such situations, higher-order contributions may lead to a bound state
where one does not exist at NLO.

%%%%%%%%%%%%%%%%
\section{The $\siii-\diii$ Coupled-Channels and the Deuteron}

The deuteron resides in the $\siii-\diii$ coupled-channels when the
$m_q$ assume their physical values.  In our previous
work~\cite{Beane:2001bc,Beane:2002vq} we found that insufficient is presently
known about the strong-interaction couplings in the two-nucleon sector to
determine if the deuteron is bound or unbound for small values of $m_q$,
including the chiral limit, or for $m_q$ larger than their physical values. In
this section we make these findings more explicit and determine the regions
allowed by data, and by NDA in the absence of data, for the $\siii$ scattering
length and deuteron binding energy, $B_d$, as a function of $m_q$, at NLO in
BBSvK-counting.

The details of the calculation of scattering lengths, phase shifts and bound
state energies in the $\siii-\diii$ coupled channels can be found in
Refs.~\cite{Beane:2001bc,Beane:2002vq}, and we do not repeat the details here.
However, a brief description of the procedure, and of the sources of
uncertainty is in order.  In this channel, unlike the $\si$-channel, the chiral
limit of one-pion-exchange (OPE) is included at LO in BBSvK-counting.  This
then allows for the re-summation of the large tensor-force that persists in the
chiral limit, and is responsible for the non-convergence of KSW-counting in
this channel~\cite{FMS}.  Deviations from the chiral limit are included in
perturbation theory, as are higher-dimension operators and multiple pion
exchanges~\cite{Beane:2001bc}.  In the $\siii-\diii$ coupled-channels, OPE
generates both central and tensor potentials,
\begin{eqnarray}
V_C^{(\pi)} (r;\mpi) & = & -{\alpha_\pi}\,\mpis\;{e^{-\mpi r}\over r} 
\quad , \quad
V_T^{(\pi)} (r;\mpi) \ =\ -{\alpha_\pi}\ {e^{-\mpi r}\over r}
\left(\ {3\over r^2} + {3 m_\pi \over r}\ +\ m_\pi^2\ \right)
\ \ ,
\label{eq:OPEpots}
\end{eqnarray}
where ${\alpha_\pi}= \gas(1-2m_\pi^2 \overline{d}_{18}/\ga)^2/(8\pi f_\pi^2)$. 
The low-energy constant (LEC) 
$\overline{d}_{18}$ has been determined through various fitting 
procedures~\cite{ulfioffe,Fettes:1999wp,Fettes:fd,Fettes:2001cr} 
to be 
$-0.78\pm 0.27$, 
$-0.83\pm 0.06$, 
$-1.30\pm 0.24$~\cite{Fettes:1999wp,Fettes:fd} and 
$-10.14\pm 0.45~{\rm GeV}^{-2}$~\cite{Fettes:2001cr}
at third-order in the chiral expansion.
While we are unable to establish a definition of the uncertainties
associated with each of these determinations, we assume that they
represent $1\sigma$ errors arising from a parabolic error
analysis of the MINUIT package, as partially described in 
Ref.~\cite{Fettes:2001cr}.
Further, in the absence of an error correlation matrix, we will treat the
uncertainties in the values of LECs as uncorrelated; this constitutes a
weakness in our propagation of errors.
Thus, we sample values of $\overline{d}_{18}$ over the range
$-1.54~{\rm GeV}^{-2} < \overline{d}_{18} < -0.51~{\rm GeV}^{-2}$,
corresponding to the $1\sigma$ limits of the first three determinations
above~\footnote{The value, $\overline{d}_{18}\sim -10~{\rm GeV}^{-2}$
~\protect\cite{Fettes:2001cr}
is significantly larger than the other determinations and we
discard it.}, and as discussed in Ref.~\cite{Fettes:fd},
this range is probably an {\it underestimate} of the true uncertainty
in  $\overline{d}_{18}$.
A fourth-order calculation has been performed and extracted values of 
$\overline{d}_{18}$ that are consistent with the three 
values above~\cite{Fettes:fd}.

As the chiral limit of the potentials in eq.~(\ref{eq:OPEpots}) 
contribute at LO, the leading $m_q$-dependence of 
$g_A$, $f_\pi$ and $M_N$ are required at NLO, as discussed in BBSvK.
Each of these observables has been studied extensively, the results of which
can be found in Refs.~\cite{ulfioffe,GaLe84,CoGaLe01,FeMe00}, and up to 
next-to-next-to-leading order (NNLO) it is known that
\begin{eqnarray}
f_\pi & = & f_\pi^{(0)} \left[ 1 
 - {1 \over 4\pi^2 (f_\pi^{(0)})^2 } 
m_\pi^2\log\left({m_\pi\over m_\pi^{{\rm PHYS}}}\right)
+ { m_\pi^2\over 8\pi^2 (f_\pi^{(0)})^2} 
\overline{l}_4  \right]
\nonumber\\
M_N & = & M_N^{(0)} - 4 m_\pi^2 c_1 
\nonumber\\
g_A & = & g_A^{(0)} 
\left[ 1 - { 2 (g_A^{(0)})^2+1
\over 4\pi^2 (f_\pi^{(0)})^2 } 
m_\pi^2\log\left({m_\pi\over m_\pi^{{\rm PHYS}}}\right)
- { (g_A^{(0)})^2\ {m_\pi^2} \over 8\pi^2 (f_\pi^{(0)})^2}
+ {4 m_\pi^2\over g_A^{(0)} } \overline{d}_{16}
\right]
\ \ \ ,
\label{eq:SNparams}
\end{eqnarray}
where $\overline{l}_4=4.4\pm 0.2$~\cite{GaLe84,CoGaLe01}, 
$c_1 \sim -1~{\rm  GeV}^{-1}$~\cite{ulfioffe} are $m_q$-independent constants
(we have explicitly separated the logarithmic contribution from 
$\overline{l}_4$) and the superscript $(0)$ denotes the chiral limit value.
We use $g_A=1.25$, $M_N=(M_n+M_p)/2$ and $f_\pi =135~{\rm MeV}$.
The parameter in eq.~(\ref{eq:SNparams})
that is most uncertain and gives rise to a substantial uncertainty
in our calculations in the two-nucleon sector 
is $\overline{d}_{16}$.  
A complete analysis by Fettes~\cite{Fettes:fd}
of the $\pi N$ sector
provides three different determinations of $\overline{d}_{16}$:
$-0.91\pm 0.74$, $-1.01\pm 0.72$ and 
$-1.76\pm 0.85~{\rm GeV}^{-2}$.
Thus we sample this parameter over the range
$-2.61 ~{\rm GeV}^{-2} < \overline{d}_{16} < -0.17$,
again corresponding to the $1\sigma$ limits of the three determinations.
In our previous paper~\cite{Beane:2002vq}, 
we retained only the chiral logarithm contribution to  $g_A$, and 
evaluated its argument  at the scale $\lambda=500~{\rm MeV}$, which corresponds
to choosing $ \overline{d}_{16}\sim +1 ~{\rm GeV}^{-2}$.
Note that while this value is outside the $1\sigma$ range of $\overline{d}_{16}$,
it is not excluded by data.
As $\overline{l}_4$ makes only  a small contribution to the 
OPE potential, its uncertainty has negligible impact and we hold 
$\overline{l}_4$ fixed to its central value in what follows.

At NLO in BBSvK-counting there is a contribution from the chiral
limit of two-pion exchange (TPE) and from an insertion of the 
$C_2$ and $D_2$ operators.
The TPE potential in coordinate space has been computed in 
Ref.~\cite{carlos,Friar:1996tj,KBW}, and in the chiral limit becomes
\begin{eqnarray}
V_C^{(\pi\pi)} (r; 0) & = & {3(22 g_A^4- 10 g_A^2-1)\over 64\pi^3 f_\pi^4} 
{1\over r^5}
\quad , \quad
V_T^{(\pi\pi)} (r; 0)  \ =\ -{15 g_A^4\over 64\pi^3 f_\pi^4} {1\over r^5}
\ \ .
\label{eq:TPEpots}
\end{eqnarray}

In order to regulate the singular potentials that occur at LO and NLO
we use a 
spatial square-well of radius $R$~\cite{Beane:2001bc,Beane:2002vq,Sprung}, 
where the 
potential outside the square well is 
\begin{eqnarray}
{\cal V}_L(r;\mpi) & = & \left(
\begin{array}{cc}   
  -M_N V_C(r;\mpi)  & -2\sqrt{2}\; M_N V_T(r;\mpi)\\               
-2\sqrt{2}\; M_N V_T(r;\mpi) & 
-M_N\left( V_C(r;\mpi)-2 V_T(r;\mpi)\right)-{6/{r^2}}
\end{array}
\right)
\ \ ,
\label{eq:potout}
\end{eqnarray}
where 
\begin{eqnarray}
V_C (r; m_\pi) & = & V_C^{(\pi)} (r; m_\pi)\ +\ 
V_C^{(\pi\pi)} (r; 0)
\quad , \quad
V_T (r; m_\pi) \ =\  V_T^{(\pi)} (r; m_\pi)\ +\ 
V_T^{(\pi\pi)} (r; 0)
\ \ .
\label{eq:potsum}
\end{eqnarray}
The potential inside the square well is
\begin{eqnarray}
{\cal V}_S(r;m_\pi ,k^2) & = & 
\left(
\begin{array}{cc}   
-M_N (V_{C_0} + V_{D_2} m_\pi^2 +k^2 V_{C_2})  & 0\\               
0     & -M_N (V_{C_0} + V_{D_2} m_\pi^2 +k^2 V_{C_2})-{6/{r^2}}
\end{array}
\right) 
\label{eq:potin}
\end{eqnarray}
where $V_{C_0}$, $V_{D_2}$ and $V_{C_2}$ are constant potentials
corresponding to the
renormalized local operators with coefficients $C_0$, $D_2$ and $C_2$
in the $\siii-\diii$ coupled-channels,
respectively~\footnote{
  The appearance of momentum-independent step-functions in 
  the D-wave potential might appear somewhat puzzling.
  However, this arises naturally when a spatial regulator 
  is used to ``smear out'' the local short-distance part of the
  central potential in a manner that is consistent with rotational
  invariance. Contributions of this D-wave component to the scattering 
  amplitude are suppressed  by powers of the square-well radius, $R$, as 
  $R\rightarrow 0$, as required in the renormalized theory.}, 
and again we have used the LO 
relation between the pion mass and $m_q$.
As discussed in BBSvK, we can make an identification between the coefficients
of local operators and the constant potentials of the square-wells
that enter into eq.~(\ref{eq:potin}), e.g.
\begin{eqnarray} 
{C_i}\ \delta^{(3)} (r) & \rightarrow & 
{{3{C_i}\  \theta (R-r)}\over 4\pi R^3}\ \equiv\  {V_{C_i}}\ \theta (R-r)  
\ \ \ ,
\label{eq:ctovsinglet}
\end{eqnarray}
and similarly for $D_i$ and $V_{D_i}$.
It is important to recall that there is implicit $m_q$-dependence in
this potential due to $g_A$, $M_N$ and $f_\pi$, in addition to the explicit
dependence from $D_2 m_\pi^2$, and from OPE.  
Defining the wavefunction $\Psi$ to be
\begin{eqnarray}
\Psi (r) & = & \left(\begin{array}{c}
            u(r)\\
            w(r)
            \end{array}\right)
          \ \ \ , 
\end{eqnarray} 
where $u(r)$ is the S-wave wavefunction and $w(r)$ is the D-wave wavefunction,
the regulated Schr\"odinger equation is 
\begin{eqnarray}
{\Psi'' (r)}
\ +\  \left[\  k^2 \ +\  {\cal V}_L(r;\mpi)\ \theta (r-R) 
\ +\ {\cal V}_S(r;m_\pi ,k^2)\ \theta (R-r)\ \right] \Psi (r) & = & 0
\ \ .
\label{eq:SE} 
\end{eqnarray} 
It is this Schr\"odinger equation that we solve to generate the scattering
length and location of the bound-state (if present) in this channel.
A detailed comparison of phase shifts generated with this formulation, and
the results of the Nijmegen partial wave analysis~\cite{Nijmegen} 
can be found in Ref.~\cite{Beane:2001bc,Beane:2002vq}.

A few words are in order concerning the power-counting.  BBSvK-counting
concerns itself with a consistent removal of cutoff effects, order-by-order in
the EFT expansion~\cite{Beane:2001bc}. The OPE potential is special in that
only the chiral limit piece requires renormalization. This is because the
$1/r^2$ contribution in the pion mass expansion vanishes and all other terms in
the expansion are regular at the origin.  The chiral
expansion of OPE has been shown to converge, albeit slowly, in Ref.~\cite{Beane:2001bc}. We
could, of course, arrange the power counting so as to keep as many orders in the
OPE perturbative chiral expansion as necessary in order to match to the results
shown in our paper, which treats the chiral expansion of OPE to all orders;
however, this would be rather pointless, since treating the potential to all
orders in the chiral expansion is perfectly innocuous from the perspective of
renormalization, and in such cases it is perfectly legitimate to treat
perturbative physics nonperturbatively. On the other hand, the TPE potential
has a significantly more divergent chiral expansion as there are
singular-attractive $1/r^5$ and $1/r^3$ contributions.  It is likely that
treating TPE to all orders while keeping only the counterterms at NLO in
Weinberg counting does not remove all cutoff artifacts (scaling as logarithms
or inverse powers of the coordinate-space cutoff). We have confirmed this strong cutoff
dependence of the full TPE numerically; this is, in fact,
strong motivation for using BBSvK counting.

\begin{figure}[!tbp]
\centerline{\psrotatefirst
\psfig{file=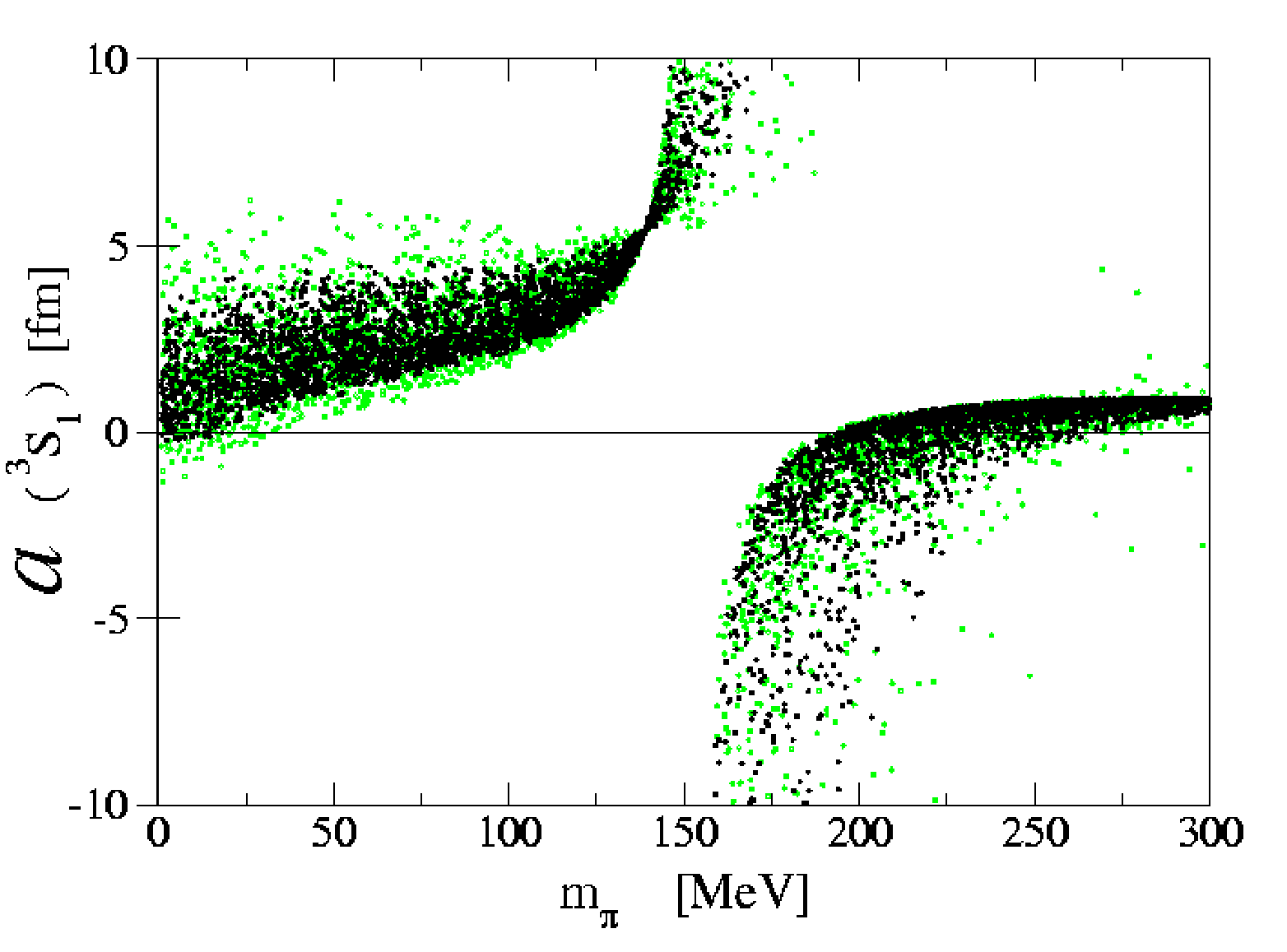,width=3.3in}}
\vskip 0.15in
\noindent
\caption{The scattering length in the $\siii$-channel as a function of the 
pion mass. The light shaded region corresponds to $\eta=1/5$
and the black region corresponds to $\eta=1/15$.
The parameter $\overline{d}_{16}$,
defined in eq.~(\protect\ref{eq:SNparams}),
is taken to be in the interval
$-2.61~{\rm GeV}^{-2} < \overline{d}_{16}
< -0.17 ~{\rm GeV}^{-2}$~\protect\cite{Fettes:fd},
and $\overline{d}_{18}=-1.54~{\rm GeV}^{-2}$.
At the physical value of the pion mass the scattering length is
$a^{(\siii)}\sim +5.425~{\rm fm}$.
}
\label{fig:3S1mqA}
\vskip .2in
\end{figure}
\begin{figure}[!tbp]
\centerline{\psrotatefirst
\psfig{file=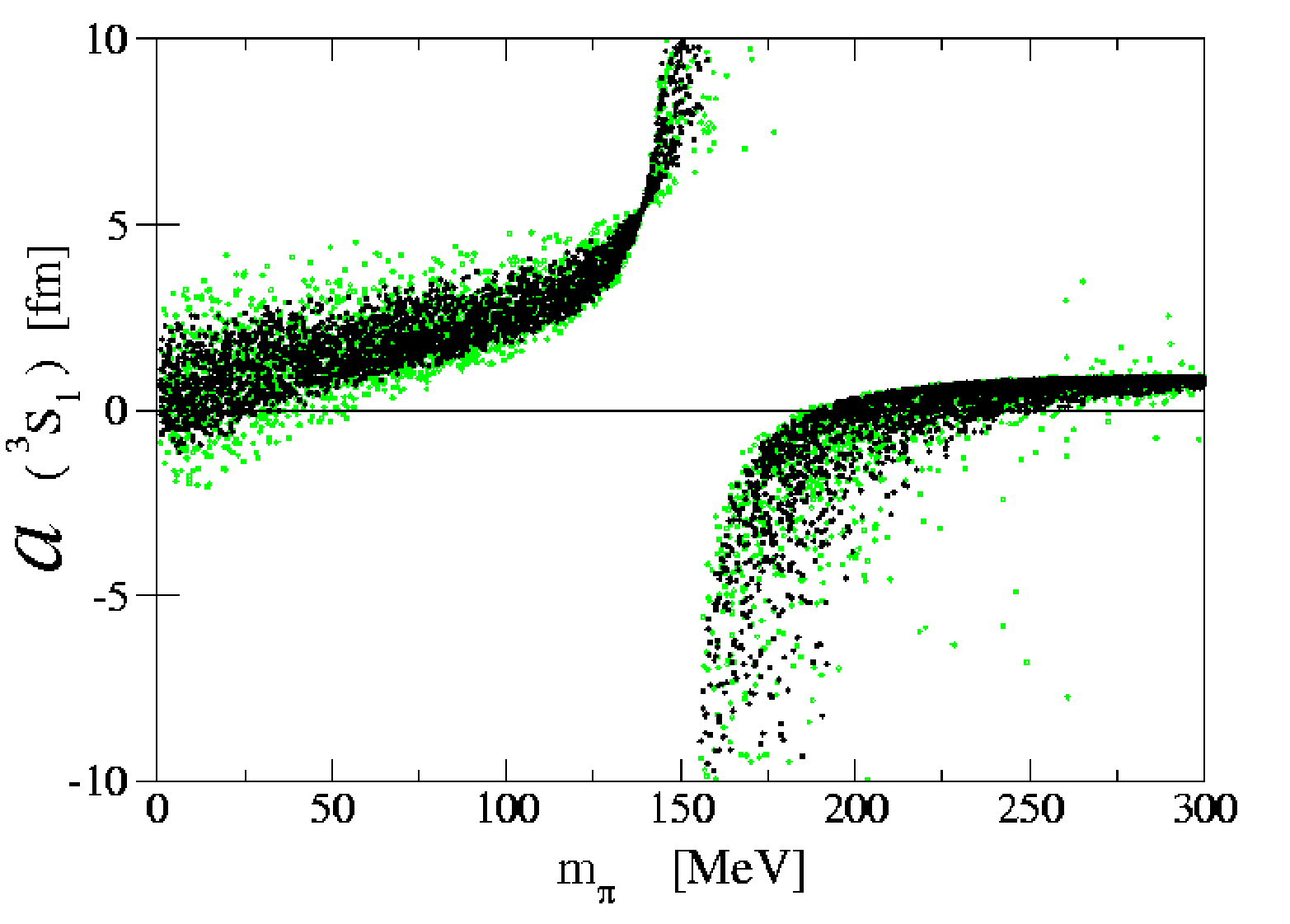,width=3.3in}}
\vskip 0.15in
\noindent
\caption{The scattering length in the $\siii$-channel as a function of the 
pion mass.
Parameters and labels are the same as Fig.~\protect\ref{fig:3S1mqA}
except $\overline{d}_{18}=-0.51~{\rm GeV}^{-2}$.
}
\label{fig:3S1mqB}
\vskip .2in
\end{figure}
Using the same definition of $\eta$ as in eq.~(\ref{eq:etadef}), we generate 
values for $D_2$ and $C_0$ consistent with NDA, and solve 
the Schr\"odinger equation in eq.~(\ref{eq:SE}) for the
scattering length in the $\siii$ channel and the deuteron binding energy.
We choose a cutoff of $R=0.45~{\rm fm}$. The cutoff-independence of our results
is discussed in Refs.~\cite{Beane:2001bc,Beane:2002vq}.
In Fig.~\ref{fig:3S1mqA}, the scattering length in the $\siii$ channel is shown
as a function of the pion mass, where we have used the LO relation between
$m_q$ and $m_\pi$. Again we have chosen $\eta$ to have the values
$\eta=1/5$ and $1/15$ to illustrate the effect of $D_2$.
We take $\overline{d}_{18}=-1.54~{\rm GeV}^{-2}$ and
$-2.61~{\rm GeV}^{-2} < \overline{d}_{16}
< -0.17 ~{\rm GeV}^{-2}$~\cite{Fettes:fd}. We again randomly and uniformly sample parameter
space within the allowed range\footnote{A more appropriate procedure might be to 
sample $\overline{d}_{16}$ from a Gaussian-weighted distribution.}.
In Fig.~\ref{fig:3S1mqB} the scattering length in the $\siii$ channel is shown
as a function of the pion mass. 
The difference with
Fig.~\ref{fig:3S1mqA} is that in Fig.~\ref{fig:3S1mqB}
we take
$\overline{d}_{18}=-0.51~{\rm GeV}^{-2}$~\footnote{While $\overline{d}_{18}$ does not
contribute in the chiral limit, it does alter the strength
of the nucleon-nucleon force at the physical value of the pion mass, where the local
four-nucleon operators are fit to data. Data, of course, contains all orders in
the chiral expansion.}.
\begin{figure}[!tbp]
\centerline{\psrotatefirst
\psfig{file=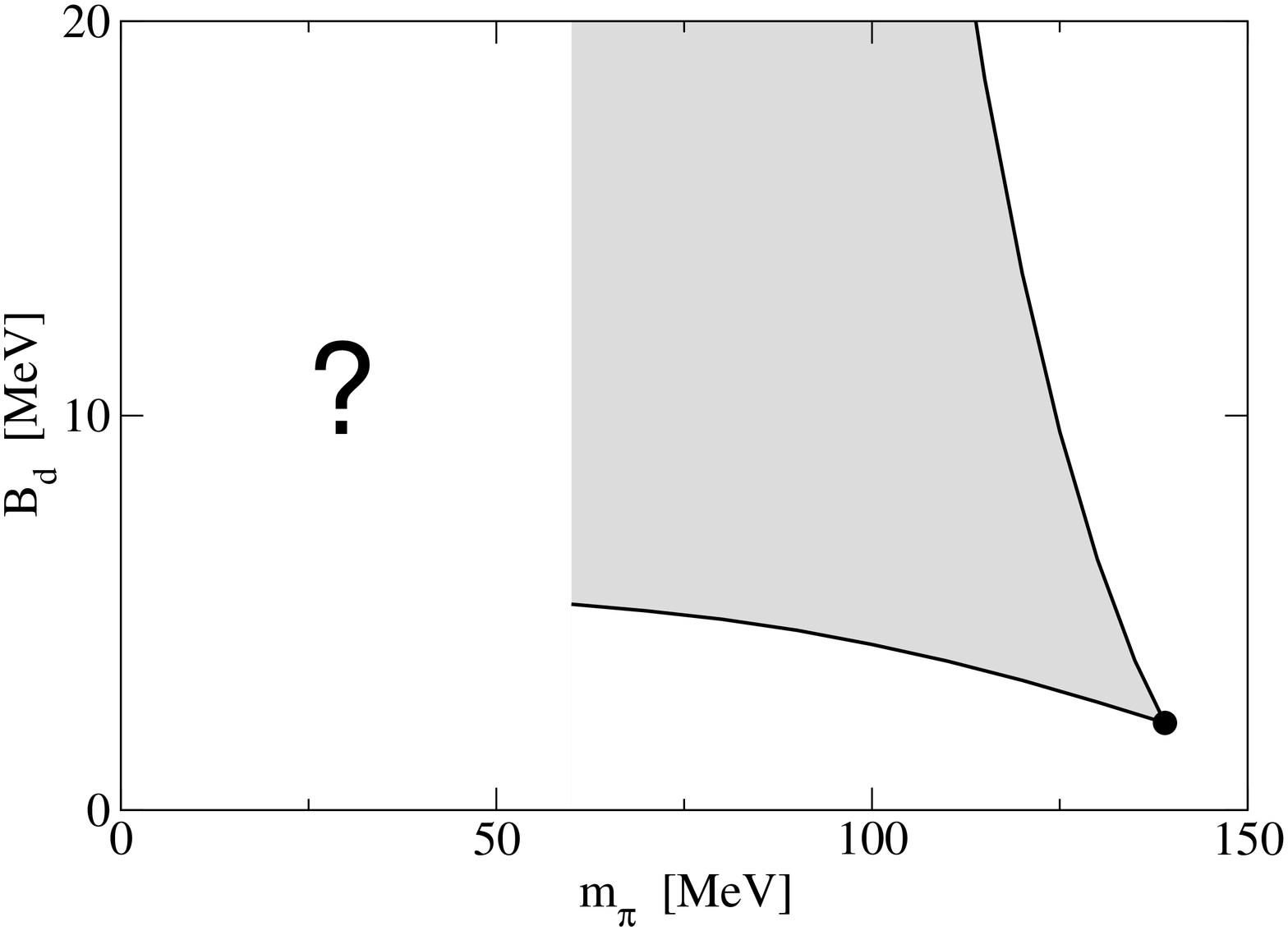,width=3.3in,angle=-90}}
\vskip 0.15in
\noindent
\caption{The deuteron binding energy as a function of the 
pion mass (from the physical value to the chiral limit). 
We do not show values below $m_\pi =60~{\rm MeV}$
since the deuteron can be both bound and unbound.
The shaded region corresponds to $\eta=1/5$, 
$-2.61~{\rm GeV}^{-2} < \overline{d}_{16} 
< -0.17 ~{\rm GeV}^{-2}$, and 
$-1.54~{\rm GeV}^{-2} < \overline{d}_{18} < -0.51 ~{\rm GeV}^{-2}$.
}
\label{fig:Bdmq}
\vskip .2in
\end{figure}
\begin{figure}[!tbp]
\centerline{\psrotatefirst
\psfig{file=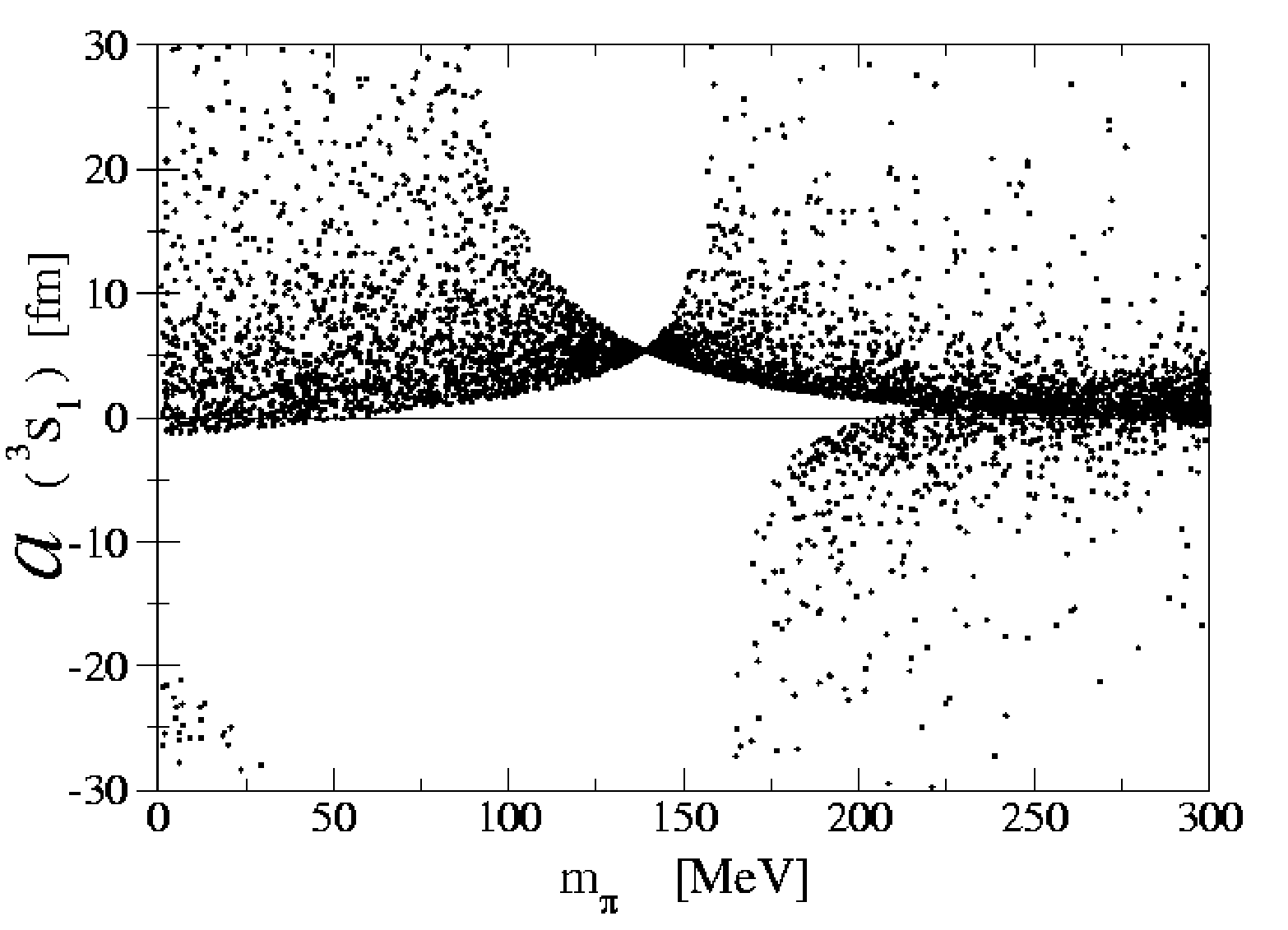,width=3.3in}}
\vskip 0.15in
\noindent
\caption{The scattering length in the $\siii$-channel as a function of the 
pion mass for $\overline{d}_{16} = +1~{\rm GeV}^{-2} $, $\eta=1/3$ and 
$\overline{d}_{18}= -0.51~{\rm GeV}^{-2}$.
}
\label{fig:3S1mqC}
\vskip .2in
\end{figure}
In Fig.~\ref{fig:Bdmq} we show the deuteron binding energy as a function of
the pion mass for the full range of
$\overline{d}_{18}$ and $\overline{d}_{16}$ and taking $\eta =1/5$.

Perhaps it is worth interpreting the binding energy curves in
Fig.~\ref{fig:Bdmq} using the scattering length curves of Fig.~\ref{fig:3S1mqA}
and Fig.~\ref{fig:3S1mqB}. It is obvious from Fig.~\ref{fig:3S1mqA} and
Fig.~\ref{fig:3S1mqB} that the scattering length can be negative for
$m_\pi\lsim 60~{\rm MeV}$, indicating an unbound deuteron.  Naively one would
expect the scattering length to go through infinity to unbind the
deuteron; this would correspond to the deuteron binding energy going through
zero. However, in this work we find that the deuteron becomes unbound in the
approach to the chiral limit by becoming increasingly deeply bound and thus
leaving the range of applicability of the EFT.

{}From these figures we conclude that the present uncertainty in 
strong-interaction parameters precludes a definitive statement being made concerning
the binding or non-binding of the deuteron in the chiral limit. This is
consistent with our previous studies~\cite{Beane:2001bc,Beane:2002vq}.
It is a combination of uncertainties in the $\pi N$ sector and the lack of any
information in the two-nucleon sector regarding  $D_2$ that conspire 
to prevent a definitive conclusion.  It is clear that progress must be made on
both issues before the chiral limit of the $\siii-\diii$ coupled channels 
can be predicted.
To demonstrate the impact of $\overline{d}_{16}$, in 
Fig.~\ref{fig:3S1mqC} we show the scattering length in the $\siii$ channel 
for $\overline{d}_{16}=+1~{\rm GeV}^{-2}$, a value that 
is not excluded by data.
Clearly the sensitivity of the two-nucleon system to this parameter is rather
dramatic. For this parameter set we see the deuteron unbinding by the
scattering length going through both infinity and zero.

Impressive progress has been made in extrapolating lattice observables in the
single-nucleon and meson sectors from lattice quark masses to the physical
quark masses.  In addition to being important in its own right, this work is
essential for interpretation of lattice QCD results in nuclei.
Extrapolations of some quenched lattice QCD observables at large quark masses
(${m_\pi}\sim 500~{\rm MeV}$) to physical values of the quark masses are seen to be in
good agreement with experimental values~\cite{bns}. Future partially-quenched or
unquenched simulations are required at smaller quark masses in order to have
confidence in first principles calculations.
As we discussed in Ref.~\cite{Beane:2001bc,Beane:2002vq}, a
quenched lattice calculation of the scattering lengths in both the $\si$ and
$\siii$-channels exists~\cite{fuku}, but for pion masses greater 
than $\sim~550~{\rm MeV}$.
While it would be tempting to extrapolate our results to these pion masses 
it is likely that the EFT will have broken down at much smaller pion masses.
Moreover, as pointed out in Ref.~\cite{BSpot}, the long-distance part of the
two-nucleon force is modified in quenched and partially-quenched QCD; for
instance, long-distance Yukawa behavior is modified to pure exponential
fall-off. This poses yet another challenge to bridging the gap between nuclear physics
and lattice QCD~\cite{BSpot}.

%%%%%%%%%%%%%%%%
\section{Comments on Recent Work by Epelbaum,\\ Gl\" ockle and
Mei\ss ner}

\begin{figure}[!tbp]
\centerline{\psrotatefirst
\psfig{file=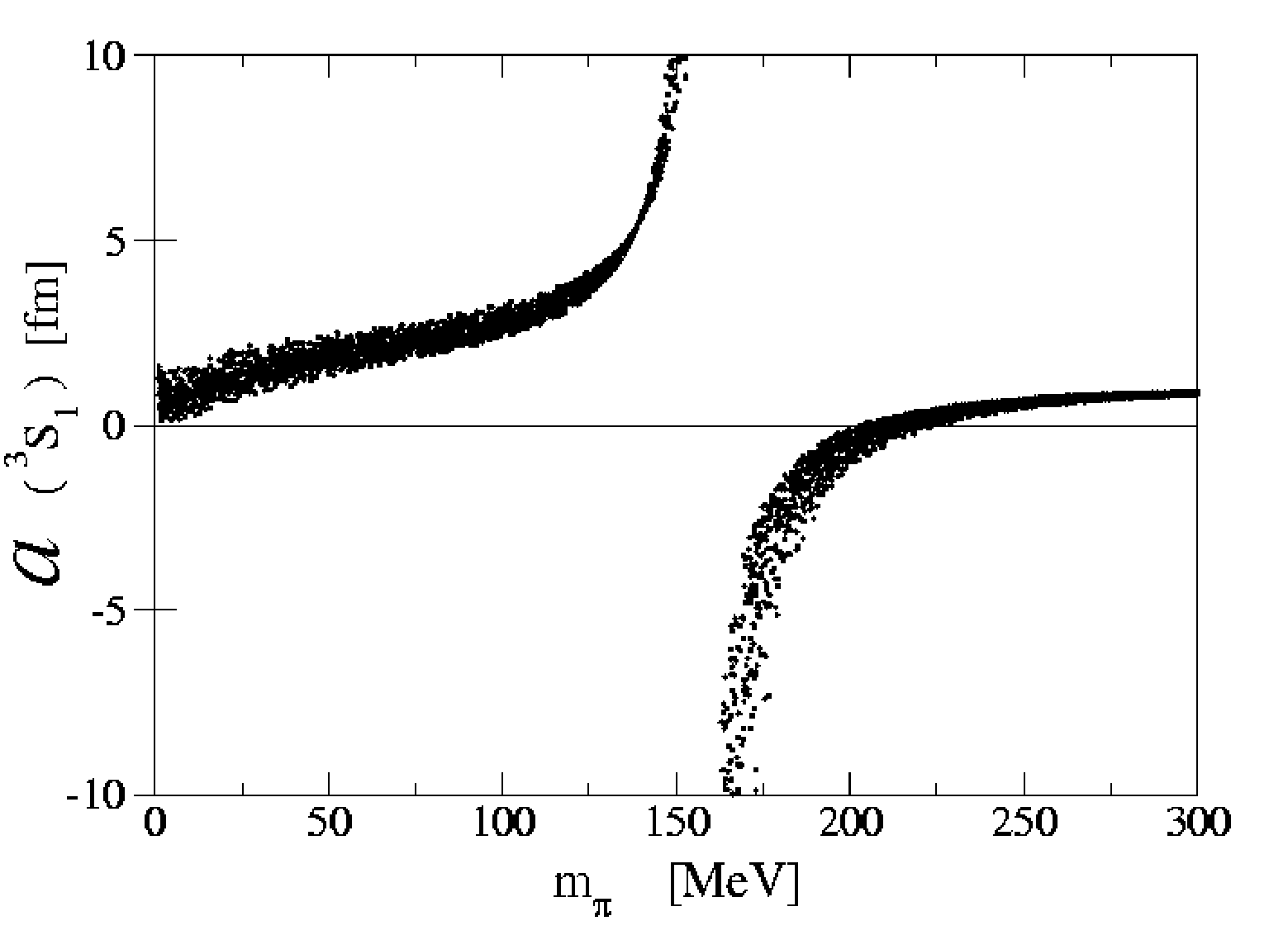,width=3.3in}}
\vskip 0.15in
\noindent
\caption{The scattering length in the $\siii$-channel as a function of the 
pion mass for the parameter set 
$-1.76~{\rm GeV}^{-2}< \overline{d}_{16} <-0.91 ~{\rm GeV}^{-2} $,
$\overline{d}_{18}=-0.97~{\rm GeV}^{-2}$ and $\eta=1/20$
used in Ref.~\protect\cite{Epelbaum:2002gb}.
}
\label{fig:3S1mqZ}
\vskip .2in
\end{figure}

While this paper was in preparation, an analysis by Epelbaum, Gl\" ockle and
Mei\ss ner (EGM) has become available~\cite{Epelbaum:2002gb} which also extends
earlier work on this subject~\cite{Beane:2001bc,Beane:2002vq}.  EGM consider
the $m_q$-dependence of nuclear forces using their formulation of Weinberg's
power-counting~\cite{We90}.  In contrast with our coordinate-space derivation,
this work involves a numerical solution to the Lippmann-Schwinger equation with
a potential generated order-by-order in the chiral expansion with a
power-counting derived for the meson and single nucleon sectors. Their work is
in qualitative agreement with the results of our previous
papers~\cite{Beane:2001bc,Beane:2002vq} and with the results we have presented
here.  Despite the differences in regularization method and the subtly
different power-counting schemes, it not surprising to see that the two EFTs
yield results that appear to be perturbatively close.  In the $\si$-channel, it
is encouraging to see that the results of the numerical calculations of
Ref.~\cite{Epelbaum:2002gb} agree with our analytic results obtained from the
perturbative-pion expansion.  However, there are quite significant differences
in the uncertainties associated with the extrapolation of the observables away
from the physical value of the pion mass, particularly in the $\siii-\diii$
coupled channels.  These discrepancies arise from a different treatment of
uncertainties associated with the strong-interaction parameters
$\overline{d}_{16}$, $\overline{d}_{18}$ and $D_2$.  It appears that
EGM~\cite{Epelbaum:2002gb} have chosen a range of values of $D_2$ consistent
with $\eta \lsim 1/19$, which we consider to be overly 
optimistic~\footnote{We should note that EGM apply their NDA argument to the two operators
$D_{S}$ and $D_{T}$. However, there are four operators,
$D_{S1,S2}$ and $D_{T1,T2}$, that can be constructed with one insertion
of $m_q$~\cite{Beane:2002vq}. With four operators present, we find that their method
saturates the range over which NLO is a subleading correction, and thus
reproduces the NDA estimates that we have used.}.
EGM argue that, for the range of cutoffs they use to define their EFT, all
experimentally determined low-energy constants in the two-nucleon sector fall
within the ``NDA'' estimates that they have created. Unfortunately, at present
there is no way to test the reliability of such estimates for $m_q$-dependent
operators.  Moreover, EGM have chosen not to propagate the known and relatively
large errors associated with the extraction of
$\overline{d}_{16}$~\cite{Fettes:fd} from the single nucleon sector.  In
Fig.~\ref{fig:3S1mqZ} we show the $\siii$ scattering length as a function of
the pion mass for the parameter ranges used by EGM.  With their parameters we
reproduce their results, up to terms higher order in both EFT descriptions. For
instance, we recover their result of $1.25~{\rm fm} \lsim a^{(\siii)} \lsim
1.85~{\rm fm}$ in the chiral limit.  Therefore, while we find the claims of
Ref.~\cite{Epelbaum:2002gb} to be quite spectacular, e.g. a deuteron binding
energy of $B_d^{(0)}=9.6\pm 1.9^{+1.8}_{-1.0}~{\rm MeV}$ in the chiral limit,
it would appear that their error analysis is incomplete, and when a complete
error analysis is performed, their claims are likely to be greatly
diminished~\footnote{An addendum to Ref.~\cite{Epelbaum:2002gb} has recently been
  archived by EGM~\cite{Epelbaum:2002erratum} which increases the error in the
  chiral limit of the deuteron binding energy by a factor $\sim 3$:
    $B_d^{(0)}=9.6^{+4.4+5.7+0.6}_{-3.2-2.4-0.4}~{\rm MeV}$.
   While their
  results are becoming more consistent with our analysis, we believe that EGM
  continue to underestimate the errors in the strong interaction parameters. For
  instance, EGM have not implemented the error correlation matrix for parameters
  in the single-nucleon sector.}.

%

%%%%%%%%%%%%%%%%
\section{Conclusions}

Understanding how nuclei and nuclear interactions depend upon the light-quark
masses is a fundamental aspect of strong-interaction physics.  We have been
able to explore the $m_q$-dependence of two-nucleon systems using a
recently-developed effective field theory and naive dimensional analysis.  In
the $\si$-channel we expect that di-nucleon systems, such as the di-neutron,
are unbound for all values of $m_q$ less than their physical values.  However,
for $m_q$ larger than their physical values both bound and unbound systems are
presently consistent with data and NDA.  In the $\siii-\diii$ coupled-channels,
where the deuteron resides for the physical values of the quark masses, the
deuteron may or may not be bound in the chiral limit.  A more definitive
statement can only be made with a more precise determination of the $\pi N$
coupling $\overline{d}_{16}$ and a determination of the coefficients of the
leading $m_q$-dependent four-nucleon operators, $D_2$. As discussed in
Ref.~\cite{Beane:2002vq}, it is likely that a determination of $D_2$ will
require a future lattice QCD calculation. While we have not investigated higher
partial waves, given that in this framework they are well described by
perturbative pion exchange, we do not expect qualitative changes to their
behavior away from the physical value of the pion mass. It is very exciting
indeed to be so close to making fundamental statements about nuclear physics.

\vskip 0.4in
%%%%%%%%%%%%%% Acknowledgements   %%%%%%%%%%%%%%%%%%%%%%
\noindent {\large\bf Acknowledgments}

\noindent
We thank the Benasque Center for Science where some of this work was performed.
We thank J.~Bjorken and U.~van Kolck for stimulating discussions and we are grateful for
useful conversations with Evgeni Epelbaum and U.-G. Mei\ss ner regarding the
low-energy constants in the single-nucleon sector.
This research was supported in part by the DOE grant DE-FG03-97ER41014.

\end{document}